      [1999/12/01 v1.4c Il Nuovo Cimento]
\documentclass{cimento}

\usepackage{graphicx}
\title{The conspicuous gamma-ray burst of 30 May 1996}
\author{F.~Ryde\from{ins:1} \atque
M.~Battelino\from{ins:1}}

\instlist{\inst{ins:1} Stockholm Observatory, AlbaNova, SE-106 91
Stockholm, Sweden}

\PACSes{\PACSit{00.00}{....} \PACSit{---.---}{\ldots}}
\begin{document}

\maketitle

\begin{abstract}
The spectra of the majority of bursts exhibit a low-energy power
law index, $\alpha$, that is either a constant or becomes softer
with time. However, in the burst of 30 May 1996 $\alpha$ becomes
harder. Here we show that this behavior can be explained by a
hybrid model consisting of a thermal and a non-thermal component.
In this burst the power-law index of the non-thermal component
changes drastically from $s \sim -1.5$ to $s \sim -0.67$ at
approximately $5$ seconds after the trigger, thereby revealing, at
low energies, the thermal component with its hard Rayleigh-Jeans
tail. This leads to the large $\alpha$-values that are found if
the Band function is fitted to the spectra. We suggest that the
change in $s$ could be due to a transition from fast to slow
cooling of the electrons emitting in the BATSE range. This could
be due to the fact that the magnetic field strength becomes
weaker.
\end{abstract}

\section{Introduction}

Despite the presence of a rich observational material and great
theoretical efforts, the prompt gamma-ray emission has defied any
simple explanation. This is in contrast to the afterglow emission
which is successfully described by synchrotron emission from a
shock moving at great speed into the surrounding medium. However,
recently Ryde~\cite{ryde} showed that the prompt emission can
indeed be described by a hybrid model of a thermal and a
non-thermal component and that the thermal component is the key
emission process determining the spectral evolution in GRBs. Even
though individual bursts appear to have complex and varying
spectral evolutions the behavior of the two separate components is
remarkably similar for all bursts, with the temperature describing
a broken power-law in time and with the non-thermal component
being consistent with emission from a population of fast cooling
electrons.

The spectral evolution found, by using the empirical Band
function~\cite{band} in the analysis, exhibits a low-energy power
law which becomes softer with time in more than half of all
cases~\cite{crider,ryde99}. Here we reanalyze the burst of 30 May
1996 which is displayed in figure 1. This burst displays the
opposite trend, that is, the low-energy power law becomes harder.
Ryde~\cite{ryde} interprets this power-law evolution as an
artifact of the empirical fit, reflecting the underlying variation
in the thermal and the non-thermal components and their relative
strengths. In most cases the thermal component is dominant during
the beginning of the evolution and thus gives rise to the
initially hard power-law fit.

\begin{figure}
\includegraphics{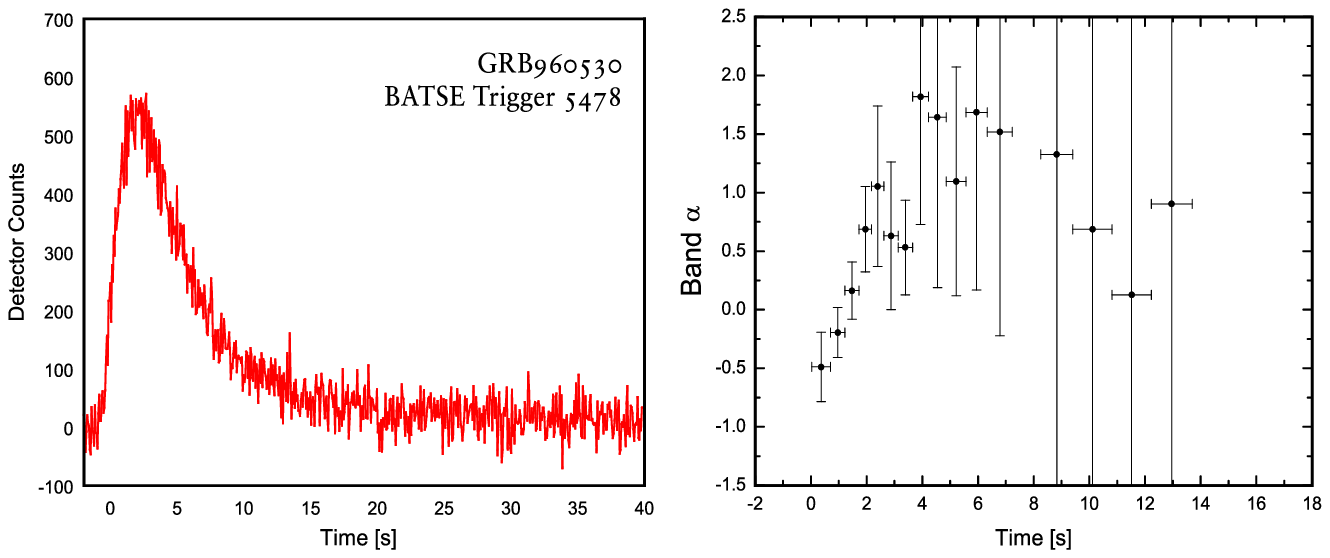}     % includes figure foo.eps
\caption{Panel a: Count light curve of the burst of 30 May 1996 as
observed by the BATSE instruments on the {\it Compton Gamma-Ray
Observatory}. Panel b: Fitting the time-resolved data with the
empirical Band function the low-energy, power-law index, $\alpha$
becomes harder at the end of the burst. This is the opposite
behavior to most bursts with strong spectral evolutions.}
\end{figure}

\begin{figure}
\includegraphics{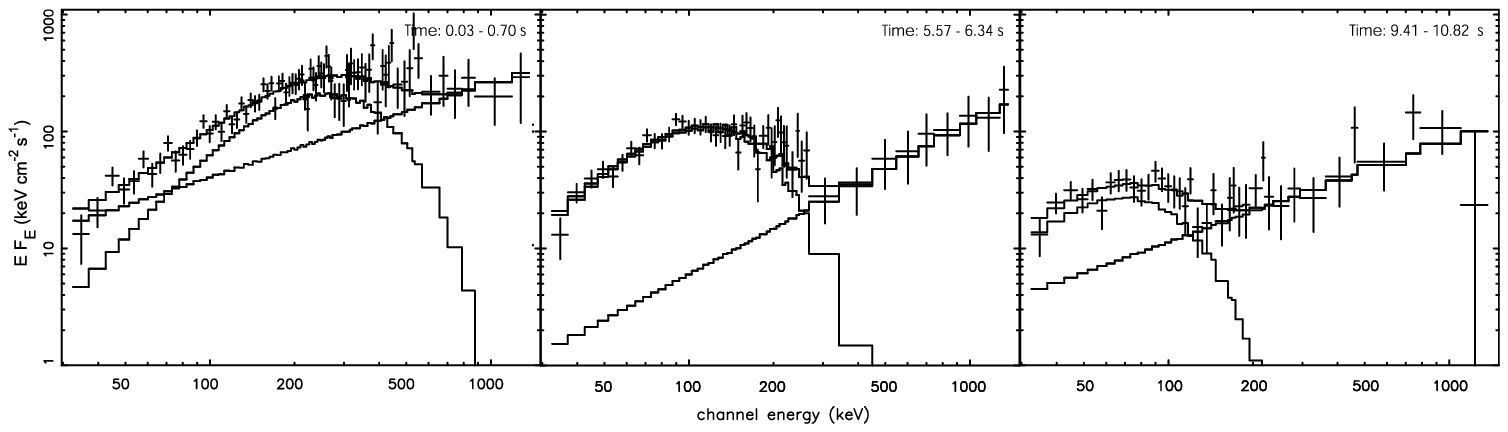}     % includes figure foo.eps
\caption{Three time-resolved spectra of GRB960530 fitted with the
hybrid model of \cite{ryde}. The time intervals that are used for
the fits are given in the upper right-hand corner of each panel.
The non-thermal spectral component is important in determining the
spectrum at the end of the pulse. }
\end{figure}

\section{Analysis}

The time-resolved spectra of GRB960530 (BATSE trigger 5478) were
modelled with the Band function~\cite{band} as well as the hybrid
model~\cite{ryde}. XSPEC version 11.3.1 was used for the spectral
analysis together with FTOOLS version 5.3.1. HERB data from the
detector with the highest rank for this burst, LAD 2, was chosen.
A background spectrum was extracted from the LAD data using
\emph{bcmppha} from the FTOOLS package. 17 spectra corresponding
to approximately the 14 first seconds after the trigger were
extracted using the same tool. Data from energy channels 1-8 and
114-128 were not included in the fitting procedure. A fit was
performed on each spectrum with the critical delta fit $10^{-5}$
for convergence. After the fitting procedure the 1$\sigma$
confidence region for each free parameter of the model was
determined. Figure 2 shows three time-resolved spectra to which
the hybrid model was fitted. In the last two spectra, the
high-energy component is important in determining the spectral
shape within the BATSE window.

To get a statistical estimate of how well a model fits the total
set of time-resolved spectra, the sum of the $\chi^2$ and the
number of degrees of freedom (d.o.f.) were calculated. The
probability that the sum of the $\chi^2$-values could exceed the
calculated $\chi^2_{\rm tot}$, if the model is correct, is then
estimated by determining the $Q$-value, $Q_{\rm tot}$. This is the
complement to the $\chi^2$-probability function. The results for
both models for the three spectra shown in figure 2 are presented
in the table below.

The hybrid model ($\chi^2_\nu = 0.975, Q=0.79$) gives a slightly
better fit to all the spectra compared to the Band model
($\chi^2_\nu = 0.999, Q=0.51$). This is in part due to the
high-energy component that is not caught by the Band model, see
figure 2. Figure 3 shows the results of these fits to the hybrid
model. The temperature decays as a broken power-law with slopes
$\sim -0.4$ and $\sim -0.7$. The thermal flux is indeed dominant
initially~\cite{ryde2}, however the power-law index, $s$, behaves
differently from other bursts \cite{ryde} in that it becomes
harder with time, thus allowing the thermal component be dominant
at low energies. We further note that $s$ makes a jump from $\sim
-1.5$ to $-0.67$ at approximately 5 seconds after the trigger.

\begin{figure}
\includegraphics{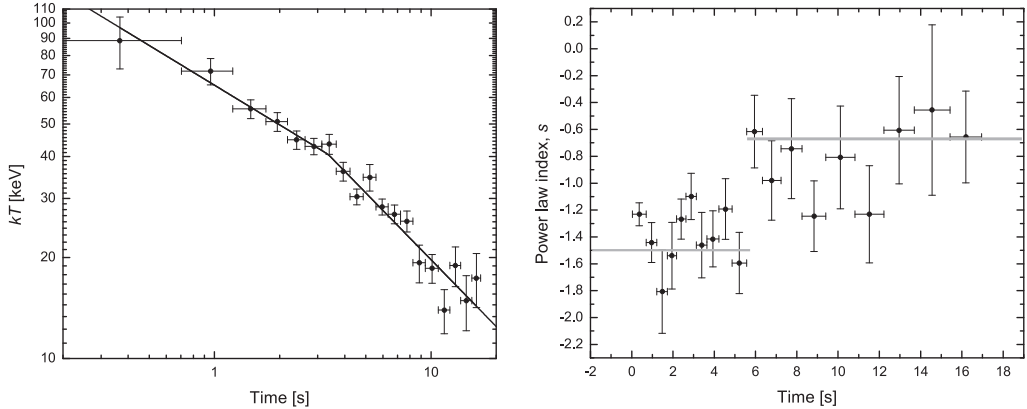}     % includes figure foo.eps
\caption{Panel a: The temperature of the thermal fit evolves as a
power law in time. Panel b: The non-thermal power-law index, $s$
makes a jump from $\sim -1.5$ to $\sim-0.67$ at approximately 5 s.
The grey lines indicate these theoretical values.}
\end{figure}

\section{Discussion}

We argue that the peculiarity of this burst, namely that  $\alpha$
becomes harder with time, is due to the fact that the non-thermal
spectral component becomes harder, letting the thermal component
dominate at low energies. This allows its Rayleigh-Jeans portion
to be exposed. The change in $s$ from $\sim-1.5$ to $\sim-2/3$ can
be interpreted as the synchrotron cooling frequency of the
electrons passing through the band-width, towards higher energies,
at around 5 s. For a population of fast cooling electrons the
spectrum below the frequency, corresponding to the minimum Lorentz
factor of the injected electrons, will have a power law index of
$-1.5$, down to the cooling frequency at which the electrons cool
on dynamical timescale. Below the cooling frequency the power-law
slope is expected to have an index of $-2/3$. Such a scenario can
be imagined if the amount of the dissipated energy going into the
magnetic fields decreases with time. The thermal component, on the
other hand, behaves similarly to other bursts with a broken
power-law decay in time (fig. 3a). It should also be noted here
that for a pure synchrotron model, instead of the hybrid model,
the observed behavior of GRB960530 would be difficult to explain.

The above investigation has shown that the hybrid
model~\cite{ryde} can give a satisfactory explanation of the
spectra and spectral evolution of GRB960530. In addition, this
burst illustrated that the non-thermal component of the hybrid
model can play an important r\^ole in determining the spectral
evolution. This component should be most important at energies
beyond the BATSE window ($\sim$ 20-2000 keV) studied here.
Furthermore, the few super-MeV detections made to date indicate
the possible presence of additional emission components at these
energies~\cite{hurley,atkins,gonz}. The GLAST satellite will be
able to address this issue in detail with its broad spectral
coverage ($\sim$ 10 keV 200 GeV) and the improved sensitivity that
it provides.

\begin{table}
  \caption{Analysis of the three time-resolved spectra shown in figure 2}
  \label{tab:1}
  \begin{tabular}{lcccccc}
    \hline
      Time interval [s]   & $\chi_{\rm hybrid}^2$ & {Q-value (hybrid)} & $\chi_{\rm band}^2$ & {Q-value (Band)} & {\rm d.o.f.}  \\
  \hline
      0.03-0.70     & 1.113&  0.205 &1.067& 0.304 &101   \\
      5.57-6.34     & 0.921 & 0.703 &1.058& 0.326 &101  \\
      9.41-10.82    & 0.887 & 0.887 &0.992 & 0.503 &101  \\
 %  \hline
 %    0.03-14 & 0.975 &  & 0.999& &  \\
   \hline
  \end{tabular}
\end{table}

\acknowledgments Support for this work was given by the Swedish
Research Council and the Swedish National Space Board.

\end{document}